\documentclass{cup-hpl}

\usepackage{lipsum}

\begin{document}

\newtheorem{theorem}{Theorem}

\shorttitle{Laboratory radiative accretion shocks on GEKKO XII}                                   
\shortauthor{L.\,Van\,Box\,Som et al.}

\title{Laboratory radiative accretion shocks on GEKKO XII laser facility for POLAR project}

\author[1,2,3]{L.\,Van\,Box\,Som\corresp{Address of corresp.
                       \email{lucile.vanboxsom@cea.fr}}}
\author[1,3]  {\'E.\,Falize}
\author[4,5]{M.\,Koenig}
\author[6]{Y.\,Sakawa}

\author[4]{B.\,Albertazzi}
\author[9]{P.\,Barroso}
\author[3]{J.-M.\,Bonnet-Bidaud}
\author[1]{C.\,Busschaert}
\author[2]{A.\,Ciardi}
\author[6]{Y.\,Hara}
\author[7]{N.\,Katsuki}
\author[6]{R.\,Kumar}
\author[4]{F.\,Lefevre}
\author[10]{C.\,Michaut}
\author[4]{Th.\,Michel}
\author[7]{T.\,Miura}
\author[7]{T.\,Morita}
\author[10]{M.\,Mouchet}
\author[4]{G.\,Rigon}
\author[6]{T.\,Sano}
\author[7]{S.\,Shiiba}
\author[6]{H.\,Shimogawara}
\author[8]{S.\,Tomiya}

\address[1]{CEA-DAM-DIF, F-91297 Arpajon, France}
\address[2]{LERMA, Sorbonne Universit\'e, Observatoire de Paris, Universit\'e PSL, CNRS, F-75005, Paris, France}
\address[3]{CEA Saclay, DSM/Irfu/Service d'Astrophysique, F-91191 Gif-sur-Yvette, France}
\address[4]{LULI - CNRS, Ecole Polytechnique,
CEA : Universit ́ Paris-Saclay ; UPMC Univ Paris 06 : Sorbonne Universit\'e - F-91128 Palaiseau Cedex, France}
\address[5]{Graduate School of Engineering, Osaka University, Suita, Osaka 565-0871, Japan}
\address[6]{Institute of Laser Engineering, Osaka University, Suita, Osaka 565-0871, Japan}
\address[7]{Faculty of Engineering Sciences, Kyushu University, 6-1 Kasuga-Koen, Kasuga, Fukuoka 816-8580, Japan}
\address[8]{Aoyamagakuin University, Japan}
\address[9]{GEPI, Observatoire de Paris, PSL Research University, CNRS, Universit\'e Paris Diderot, Sorbonne Paris Cit\'e, F-75014 Paris, France}
\address[10]{LUTH, Observatoire de Paris, PSL Research University, CNRS, Universit\'e Paris Diderot, Sorbonne Paris Cit\'e, F-92195 Meudon, France}

\begin{abstract}
A new target design is presented to model high-energy radiative accretion shocks in polars. In this paper, we present the experimental results obtained on the GEKKO XII laser facility for the POLAR project. The experimental results are compared with 2D FCI2 simulations to characterize the dynamics and the structure of plasma flow before and after the collision. The good agreement between simulations and experimental data confirm the formation of a reverse shock where cooling losses start modifying the post-shock region. With the multi-material structure of the target, a hydrodynamic collimation is exhibited and a radiative structure coupled with the reverse shock is highlighted in both experimental data and simulations. The flexibility on the laser energy produced on GEKKO XII, allowed us to produce high-velocity flows and study new and interesting radiation hydrodynamic regimes between those obtained on the LULI2000 and Orion laser facilities. 
\end{abstract}


\maketitle

\section{Introduction}
Accretion processes are main sources of high-energy radiations in several binary systems \citep{Franck2002}, in particular in cataclysmic variables. As potential progenitors of type Ia supernovae \citep{Maoz2014}, understanding these complex systems is crucial to explain the initial conditions of these explosions which are used to study the acceleration of the Universe \citep{Riess1998}. Among these objects, polars are remarkable ones to study accretion processes in isolation since the high-energy radiation coming from accretion processes are not contaminated by other surrounding luminosities. They are close binary systems composed of a strongly magnetized white dwarf accreting matter from a low-mass companion star \citep{Warner1995}. The intense magnetic field of the white dwarf ($B_{\text{WD}}> 10$\,MG) locks the whole system into synchronous rotation, prevents the formation of accretion disks and guides the accreted flow as an accretion column onto the white dwarf magnetic poles \citep{Cropper1990, Wu2000}. The supersonic accreted flow coming from the companion and channelled by the dipolar magnetic field, strikes the white dwarf photosphere at the free-fall velocity ($v_{\text{ff}}\sim 3000$\,km\,s$^{-1}$), creating an accretion shock. This shock is counter propagating counter to the incoming flow and heats the matter to temperatures of about $10$\,keV. Thus the intense emitted radiation shapes the post-shock region and slows down the accretion shock which reaches a height of about $100$\,km above the white dwarf photosphere. Consequently, the small size of this post-shock region prevents direct inference of spatial profiles of the radiative zone which are relevant to determine the white dwarf mass. Besides, the observed flux can not be reconciled with the current standard model of accretion columns which introduce strong disagreements between theory and observations \citep{Ramsay2004, Bonnet2015, Mouchet2017,VBS2018}. Thus, powerful facilities provide an alternative approach which will help clarify outstanding questions related to radiative processes in accretion column models.

Based on similarity properties of this high-energy environment, millimetre-sized models of accretion columns can be built with powerful lasers. The radiation hydrodynamics physics of these structures admits exact scaling laws which demonstrate that measurable-scale models of radiative accretion columns could be produced in laboratory for the main accretion shock regimes: the bremsstrahlung-dominant regime \citep{Falize2009_scaling}, the two radiative processes regime (bremsstrahlung and cyclotron processes) \citep{Falize2011_scaling}, the two-temperature regime \citep{Falize2011_2T}, and the ideal magnetohydrodynamical (MHD) regime \citep{Ryutov2000}. These scaling laws for many different shock regimes offer new opportunities to study astrophysical objects at laboratory scale. To assure the relevance of such experiments, the key dimensionless numbers of the laboratory and astrophysical plasmas have to be equal which implies that the two physical regimes are similar. In this case, powerful lasers become microscopes to study radiative processes at laboratory scale relevant to the astrophysical one.

Based on the Alfven similarity \citep{Ryutov2000}, the calculated magnetic field at laboratory scale is given by $B_{\text{L}} = \sqrt{P_{\text{L}}/P_{\text{A}}} B_{\text{A}}$ where $P_{\text{L}}/P_{\text{A}}$ is the ratio of the laboratory and astrophysical pressures and $B_{\text{A}}$ is the intensity of the astrophysical magnetic field. For typical values in polars,  the laboratory magnetic field should be equal to about $1$\,GG. This intensity of magnetic field given by the scaling laws is obviously inaccessible in the laboratory, except in very particular configurations \citep{Korneev2015}. This prevents from studying the main polars regime with the multiple radiative processes. Consequently, the only accessible accretion regime with lasers is the bremsstrahlung-dominant regime when the white dwarf magnetic field is expected to be weak ($B_{\text{WD}} = 10-30$\,MG). In this regime, the only magnetic effect is the magnetic collimation of the column due to the white dwarf magnetic field and others magnetic effects, such as the cyclotron radiation, can be neglected \citep{Warner1995}. At laboratory scale, to replace magnetic collimation, a tube was used to mechanically collimate the flow. This design has been successfully experimented on LULI2000 \citep{Falize2011_POLAR,Falize2012} and Orion \citep{Cross2016} laser facilities in the POLAR project \citep{Falize2011_POLAR} and simultaneously on the OMEGA facility \citep{Krauland2013, Krauland2013b} laser facility. In spite of the success of these experiments and their relevance for the study of high-energy processes, the plasma flow collimation using the tube can generate various wall shocks before impact and afterwards, it can corrupt the reverse shock formation and propagation. Besides, to implement X-ray radiography diagnostics, holes are introduced in the tube which induce 3D effects on the reverse shock structure. Consequently a design is therefore under study to remove the tube where the collimation of the supersonic flow is produced by a dynamical nozzle-like effect.

In this paper, experimental results, obtained with the GEKKO XII laser facility with a new target design, are presented and compared to 2D FCI2 laser simulations. Many optical diagnostics have been implemented to probe the dynamics and the structure of the plasma flow and the post-shock region. In the first section, the target design is presented. In the second one, the dynamics of the incident flow and the properties of the post-shock region are presented through experimental data and 2D simulations. Finally, the similarity properties of the target are investigated with the GEKKO XII laser conditions.

\section{Target design and experimental setup}

\begin{figure}[h!]
\centering
\includegraphics[width=9cm]{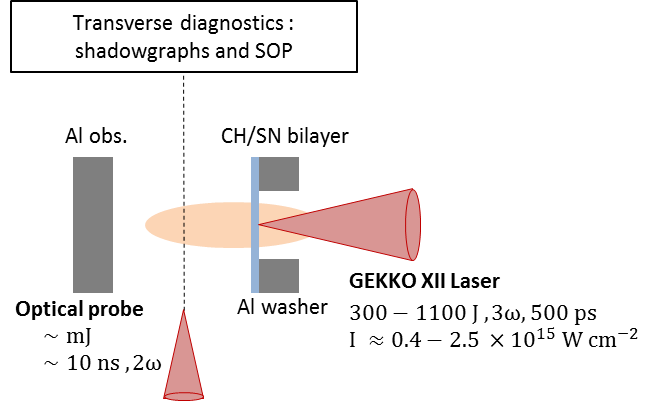}
\caption{Target's schematic. The laser is coming from the right and it interacts with the pusher. A plasma is created due to the interaction between the laser and the pusher. This supersonic plasma expands in the vacuum and impacts an obstacle. This leads to the creation of a reverse shock. }\label{cible}
\end{figure}

The target design, which models the astrophysical scale, is shown in Figure\,\ref{cible}. It is composed of a pusher playing the role of the accreted matter, and an obstacle modelling the white dwarf photosphere. The two parts are separated by a vacuum zone and fixed on a plastic support. The main difference from previous POLAR targets is the absence of the tube.

The pusher is composed of a $25$\,$\mu$m thickness plastic layer ($1.29$\,g\,cm$^{-3}$), which converts laser energy into kinetic energy by rocket effect \cite{Atzeni}, and a $4$\,$\mu$m thickness tin layer, which has two important roles: it protects the obstacle against X-ray radiation produced by the hot coronal plasma and increases the radiative properties of the target. All these layers are stuck to an aluminium washer. 

Experiments were performed on the GEKKO XII facility. Depending of the chosen laser configuration, the energy delivered on target can range from $300$\,J to $1100$\,J at $3\omega$ wavelength, with a Gaussian focal spot of around $350-400$\,$\mu$m FWHM and a Gaussian pulse width of $500$\,ps. The associated intensities can vary from $4 \times 10^{14}$\,W\,cm$^{-2}$ to $2.5 \times 10^{15}$\,W\,cm$^{-2}$ on target. Seven optical diagnostics (1D and 2D SOP, 1D and 2D shadowgraphies and interferometry) have been implemented to probe the incoming plasma and the post-shock region. An optical probe beam ($\sim$mJ, $2\omega$ and $10$\,ns duration) has been installed in the perpendicular direction to the incoming flow propagation direction. Transverse optical pyrometry (SOP) diagnostics image the self-emission from the incoming flow and the reverse shock structure. In addition, shadowgraphy diagnostics record the global shape of the plasma and interferometers enable inference of the electron density of the flow. Streak cameras record the incoming flow expansion and propagation in vacuum, and the radiation flux emission as a function of time. 

We compare experimental results with one-dimensional (1D) and two-dimensional (2D) numerical simulations performed with the CEA laser radiation hydrodynamic Arbitrary-Lagrangian-Eulerian (ALE scheme) code FCI2 \citep{Schurtz2000}. We use the multi-group diffusion model ($100$ groups) which allows one to reproduce the GEKKO XII regime and the laser-matter interaction is modelled by a ray tracing algorithm \cite{Busschaert2013}. The typical longitudinal and radial resolutions of the simulations are around $0.5$\,$\mu$m.

\begin{figure}[h!]
\centering
\includegraphics[width=8cm]{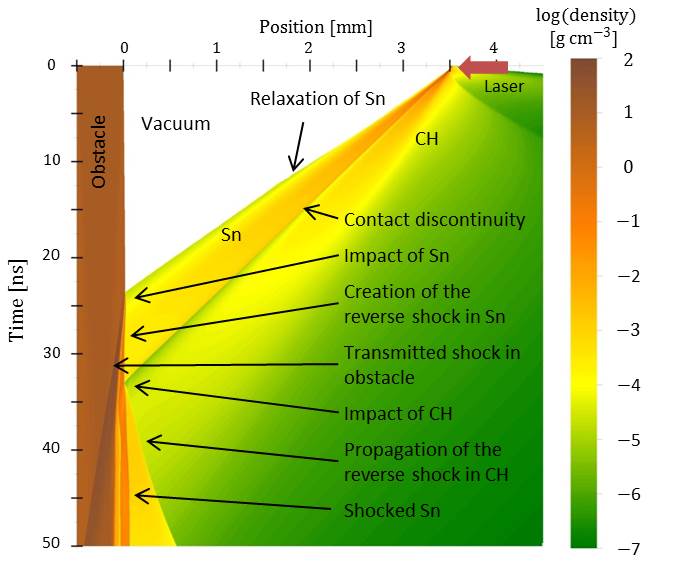}
\caption{Spatial evolution of the density as function of time extracted from 1D simulation performed with FCI2 code. The position axis is horizontal whereas the time evolution is the vertical axis. The laser deposits $600$\,J to the target and it is coming from the right. The distance to the obstacle is $3.5$\,mm.}\label{1d_bi}
\end{figure}

To illustrate the experimental principle and the dynamics of the plasma flow, a 1D numerical simulation result is shown in Figure\,\ref{1d_bi}. We present the density evolution as function of time. The plasma, created by the laser-pusher interaction, expands in the vacuum. Then, the supersonic plasma flow impacts an aluminium solid obstacle leading to the formation of a radiative reverse shock which can be analysed with laser diagnostics. At the same time, a transmitted shock is created in the obstacle. The radiative properties of the post-shock region are proportional to the charge number $Z$ \citep{Busschaert2013}, which justifies to use a high-$Z$ layer, here the tin layer, to increase radiative effects. Despite the obvious simplicity of the target design, the studied physical processes (collision of an expanding plasma flow with an obstacle and formation of a radiative reverse shock) are relatively hydrodynamical complex issues where high-energy radiation can play a fundamental role.

To study the density and the temperature evolutions in the vacuum, two distances between the pusher and the obstacle, labelled $l_{\text{vac}}$, are considered: $2.5$\,mm and $3.5$\,mm. Indeed, the density ($\rho_{\text{flow}}\propto l_{\text{vac}}^{-1}$) and the temperature ($T_{\text{flow}}\propto [l_{\text{vac}}]^{1-\gamma}$ considering an adiabatic flow) of the expanding plasma are inversely proportional to powers of the distance to travel. 

\section{Laboratory accretion plasma}\label{section_simu} 

\subsection{Plasma flow generation and propagation}

\begin{figure}[h!]
\centering
\includegraphics[width=7cm]{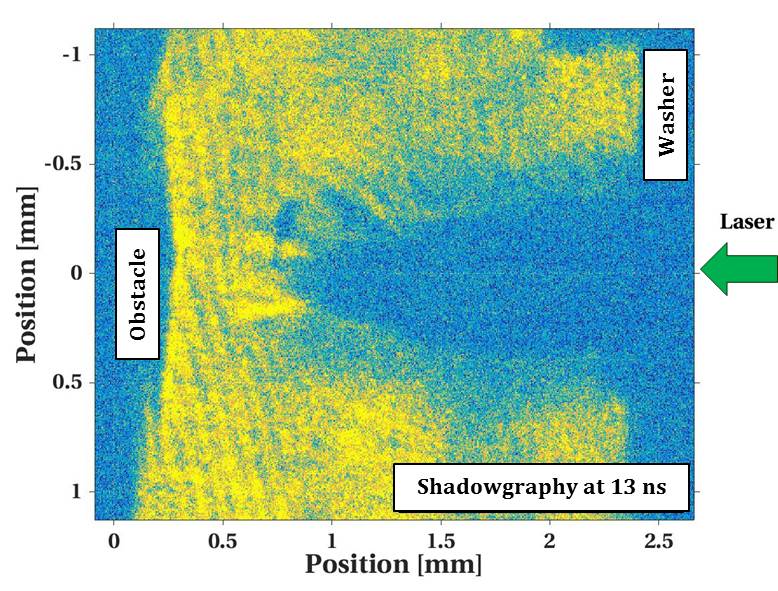}
\caption{2D snapshot shadowgraphy obtained at $13$\,ns after the laser drive of the incident plasma flow. }\label{fig_shadow_2d}
\end{figure}

First, it is necessary to characterize the plasma structure and dynamics before the collision since the reverse shock depends on the physical properties of the incident flow. A typical image from shadowgraphy diagnostics is presented in Figure\,\ref{fig_shadow_2d}. The length of the flow is $1.6$\,mm around $7$ ns before the collision and its radial expansion is $860$\,$\mu$m for a laser energy of $836$\,J. 

\begin{figure}[h!]
\centering
\includegraphics[width=8.5cm]{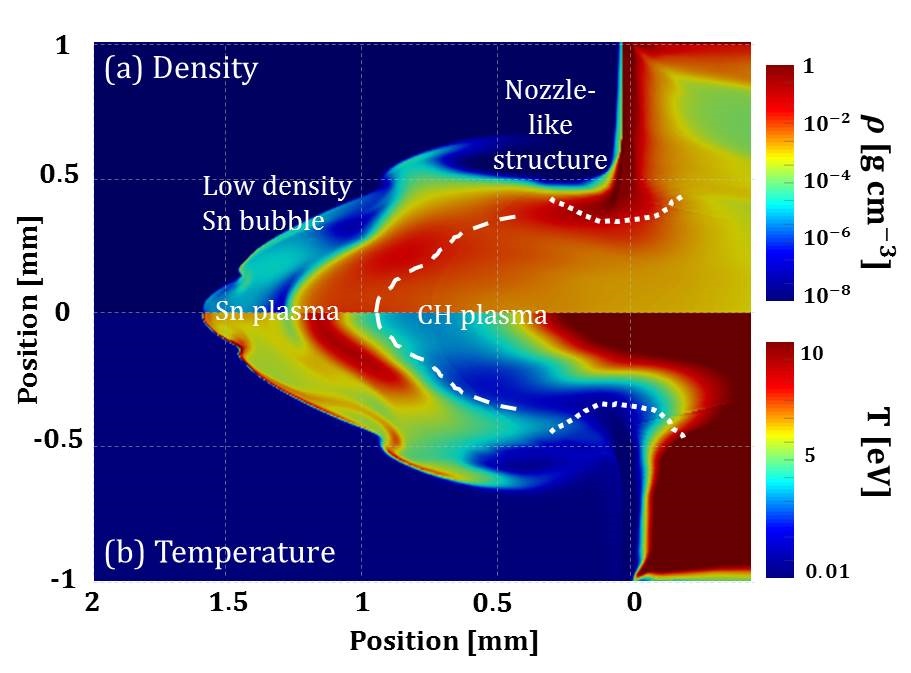}
\caption{Density (a) and temperature (b) maps of the incident flow around $10$ ns before the collision extracted from 2D FCI2 simulation.}\label{fig_simu1}
\end{figure}

We compare these values with a 2D axi-symmetrical numerical simulation. The density and temperature maps of the incident flow around $10$ ns before the collision obtained by simulation are presented in Figures \ref{fig_simu1} (a and b). Due to the absence of tube, the Sn plasma expands in vacuum by creating a bubble around the CH plasma. The contact discontinuity between Sn and CH layers can be observed in Figure \ref{fig_simu1}b that confirms the bubble structure of the flow. The velocity of the high-$Z$ radial bubble expansion can be evaluated to about $20$\,km\,s$^{-1}$ which corresponds to the transverse kinetic energy lost for the propagation. The dimensions of the flow obtained by simulations (Figure \ref{fig_simu1}a) are compatible with those of the experimental data in Figure \ref{fig_shadow_2d}. Moreover a collimation effect is induced by Sn expanding plasma stuck on the washer which creates important pressure gradient located at the internal radius of the aluminium washer. The expanding plasma is due to the propagation of thermal waves created by the interaction laser with matter, where electrons transport energy. This structure generates an analogue nozzle geometry. This effect induces a hydrodynamic collimation of the flow visible both in experimental data (see Figure \ref{fig_shadow_2d}) and simulation (see Figure \ref{fig_simu1}).

\begin{figure}[h!]
\centering
\includegraphics[width=9cm]{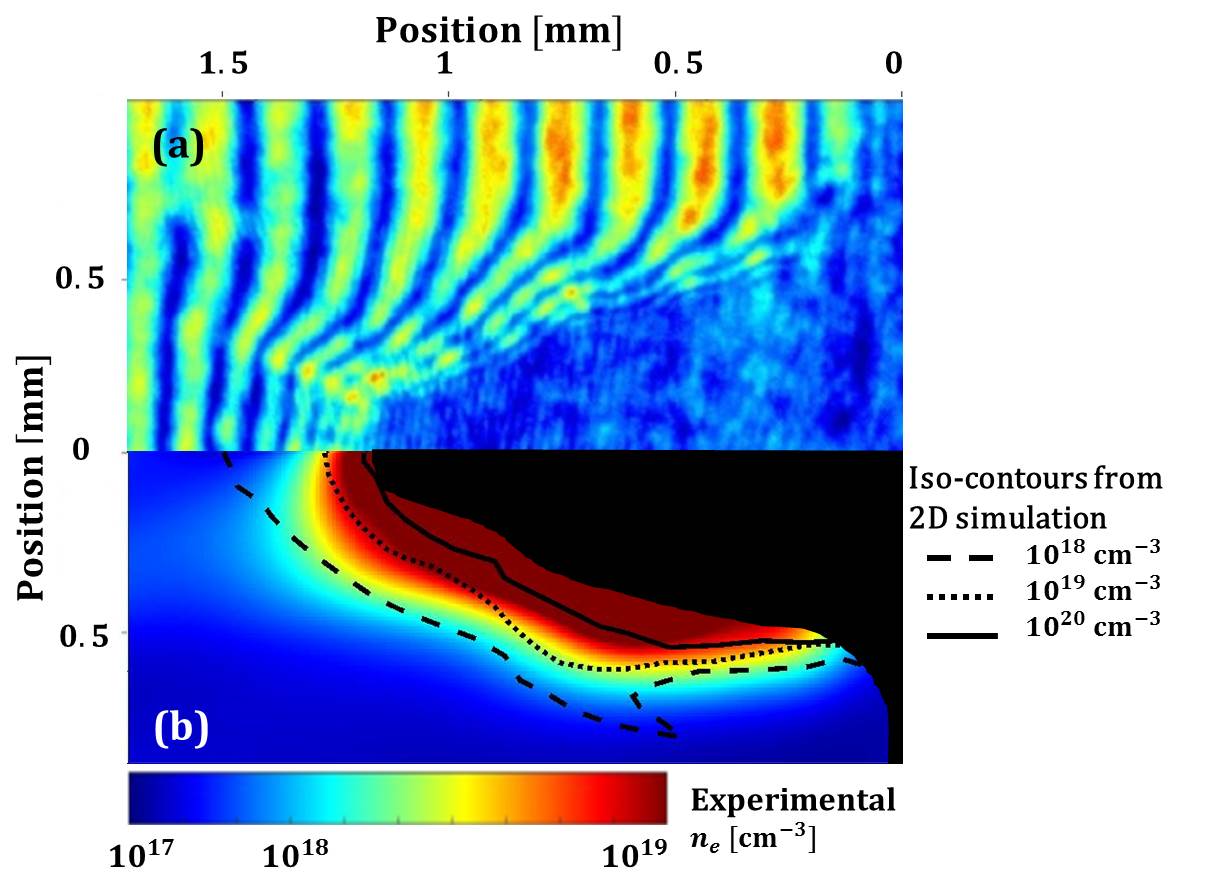}
\caption{2D snapshot interferometry (a) obtained at $11$\,ns after the laser drive of the incident flow compared with the associated electronic density (b). The experimental electronic density is compared to iso-density curves at $10^{18}$, $10^{19}$ and $10^{20}$\,cm$^{-3}$ extracted from 2D simulations (black lines).}\label{fig_density}
\end{figure}

In the shadowgraphy, a low density plasma is observed. It is confirmed in the 2D snapshot interferometry (Figure\,\ref{fig_density}a) where we can observe strong perturbations of the fringes around the plasma flow. These irregularities are due to lower plasma expanded ahead the flow. The electronic density of this low-density matter is presented in Figure\,\ref{fig_density}b. It has been calculated assuming a cylindrical symmetry using Neutrino software\footnote{GitHub repository https://github.com/NeutrinoToolkit/Neutrino}. The iso-density curves at $10^{18}$, $10^{19}$ and $10^{20}$\,cm$^{-3}$ extracted from 2D simulations are added to the experimental electronic density (see black lines in Figure \ref{fig_density}b). In spite of some disparities due to large errors in the determination of the fringe shifts, a global agreement is found between the experimental and numerical plasma shape.

\begin{figure}[h!]
\centering
\includegraphics[width=8cm]{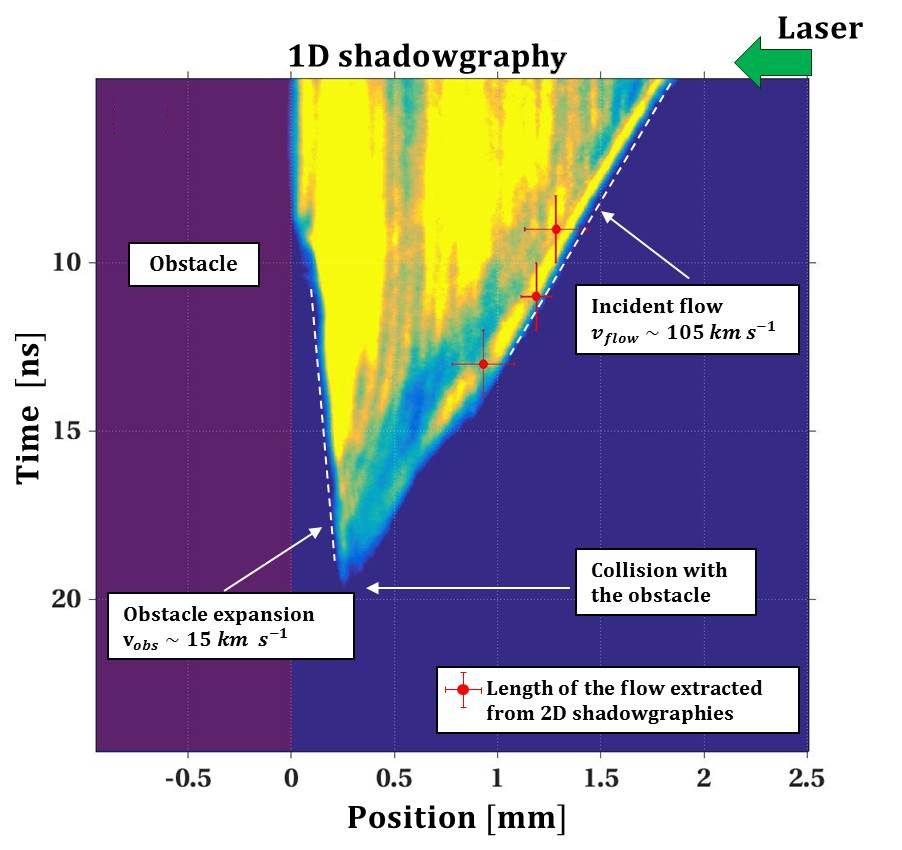}
\caption{1D shadowgraphy used to diagnose the velocity of the incident flow. The plasma created by the laser is coming from the right. The position is relative to the obstacle position whereas the vertical axis presents the time evolution. }\label{fig_40376_1d}
\end{figure}

\begin{figure}[h!]
\centering
\includegraphics[width=8.8cm]{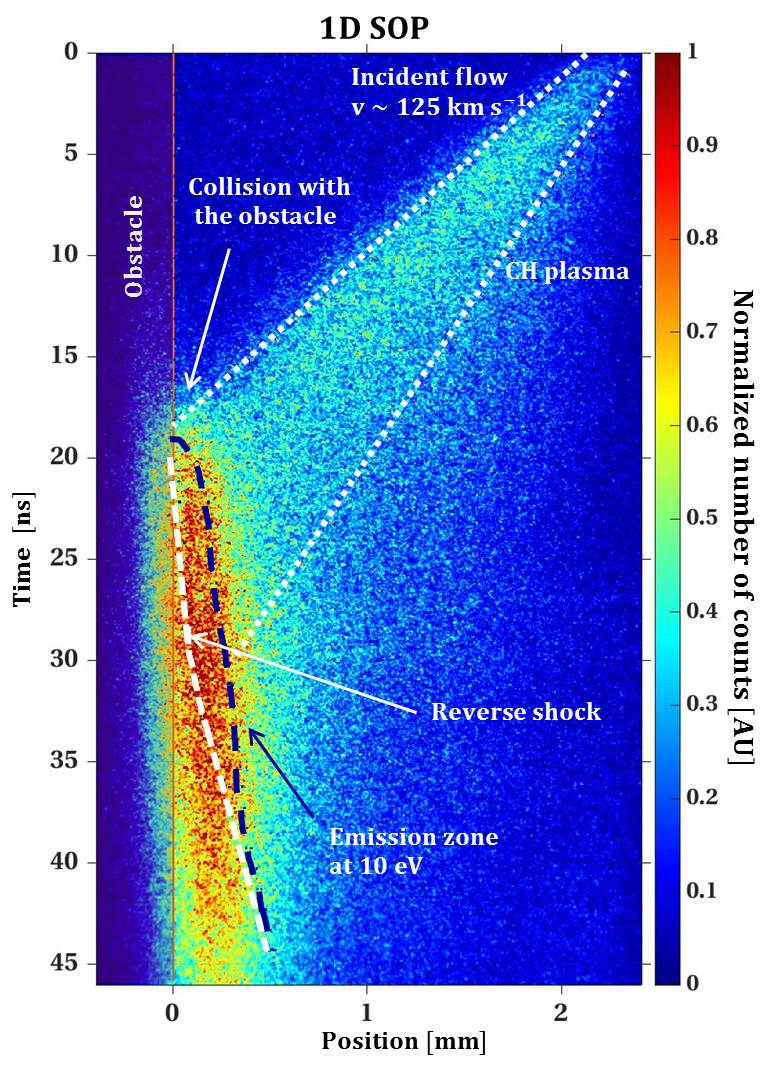}
\caption{1D self-emission (SOP). The plasma created by the laser is coming from the right. The position is relative to the obstacle position whereas the vertical axis presents the time evolution. The CH flow position (white dotted line), the reverse shock position (white line) and the $10$\,eV iso-contour (black line) extacted from 2D simulations are added.}\label{fig_40376_1d_sop}
\end{figure}

From 1D transverse shadwography and 1D self-emission diagnostic, we can infer the expansion velocity that is prime importance to the radiative properties of the post-shock region as we will see later. Typical images are presented respectively in Figures\,\ref{fig_40376_1d} and \ref{fig_40376_1d_sop}. The plasma created by the laser is coming from the right. The position is relative to the obstacle position whereas the vertical axis presents the time evolution. The mean velocity of the incoming flow is $115$\,km\,s$^{-1}$ with a collision time at about $20$\,ns. 

We can notice that the incoming flow position is slightly overestimated with the SOP diagnostic ($v_{\text{flow}}= 125$\,km\,s$^{-1}$) compared to the shadowgraphy ($v_{\text{flow}}= 105$\,km\,s$^{-1}$). The plasma density, probed by the shadowgraphies, is close to the critical density, $n_{e}\sim n_{c} \sim 10^{21}$\,cm$^{-3}$ whereas the radiation is emitted by a lower density zone with $n_{e}<10^{20}$\,cm$^{-3}$ in the front of the flow. The flow length determined with 2D shadowgraphy snapshots are added on the 1D shadowgraphy presented in Figure\,\ref{fig_40376_1d}a (red points). Good agreement is found between the 1D and 2D shadowgraphies. In the 2D FCI2 simulations the plasma flow at lower density and at about $5$\,eV propagates at $\sim 130$\,km\,s$^{-1}$ whereas at about $n_{c}$, it propagates at a lower velocity, $\sim 100$\,km\,s$^{-1}$. Good agreement is found for the impact time and for the velocity of the plasma flow. The simulation forecasts a hypersonic incident flow with an internal Mach number around $14$. During the expansion of the incident flow, the obstacle is heated before the collision due to the high-energy radiation coming from the coronal plasma. The expansion velocity of the obstacle is about $15$\,km\,s$^{-1}$ as shown in Figure \ref{fig_40376_1d}a. This does not disturb the collision and the generation of the reverse shock. 

\begin{figure}[h!]
\centering
\includegraphics[width=8cm]{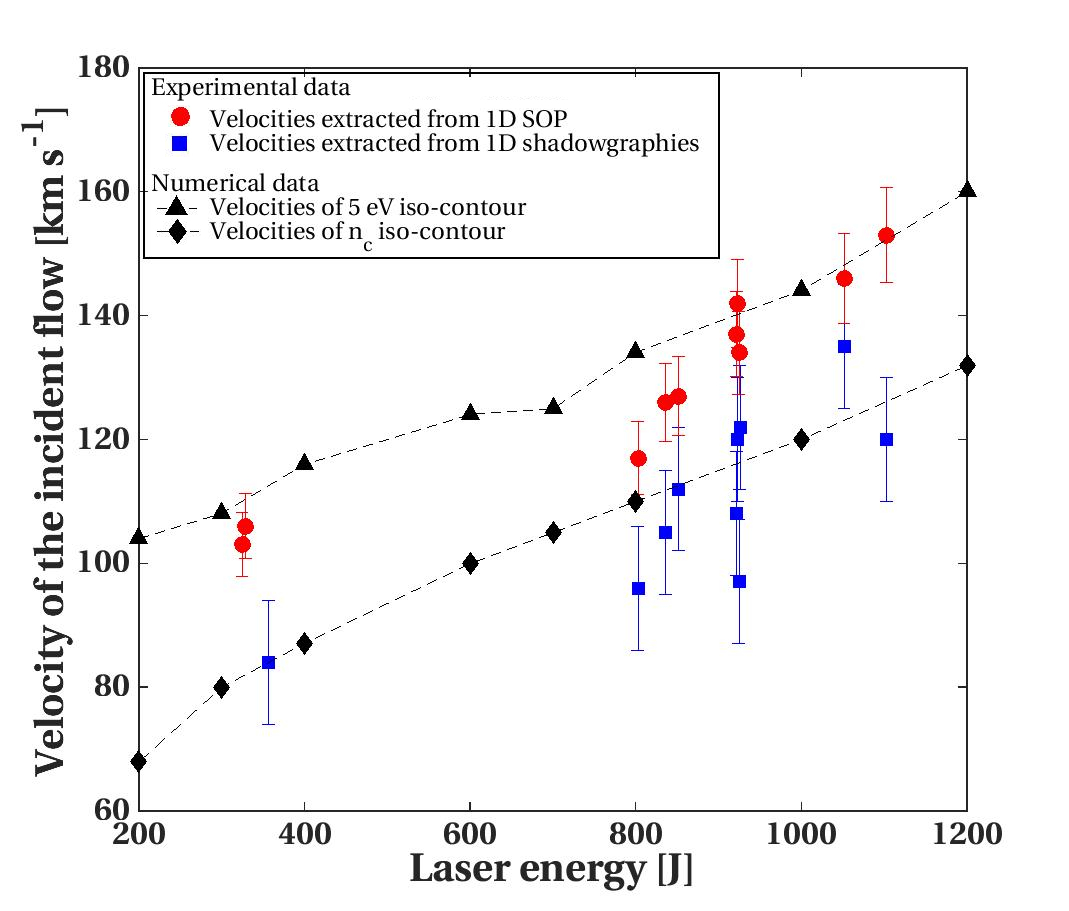}
\caption{Experimental velocities of the incident flow as function of the laser energy extracted from the 1D shadowgraphies (blue squares) and from the 1D SOP (red points). They are compared to velocities extracted from 2D simulations : velocities of iso-temperature curve at $5$\,eV (black triangles) and velocities of the iso-density curve at $n_{c}$ density (black diamonds).}\label{fig_velocities}
\end{figure}

Figure\,\ref{fig_velocities} summarizes the different plasma flow velocities determined from the experimental data as a function of the laser energy achieved on the GEKKO XII facility. The errors are associated to the uncertainties in the plasma position determination from the diagnostics which are linked to the uncertainties of the plasma  velocities. We compared the plasma flow velocities extracted from 2D simulations (black triangles and diamonds) to the 1D shadowgraphy (blue squares) and the 1D SOP (red points). The SOP velocities (red points) are compared to velocities of the iso-temperature curve at $5$\,eV in the incident flow at different laser energies (black triangles). Then the shadowgraphy velocities (blue points) are correlated with the velocities of iso-density curve at $n_{c}$ density (black diamonds). A  good tendency is obtained between experimental and numerical data for the two types of flow.

\subsection{Reverse shock structure}

\begin{figure}[h!]
\centering
\includegraphics[width=8.5cm]{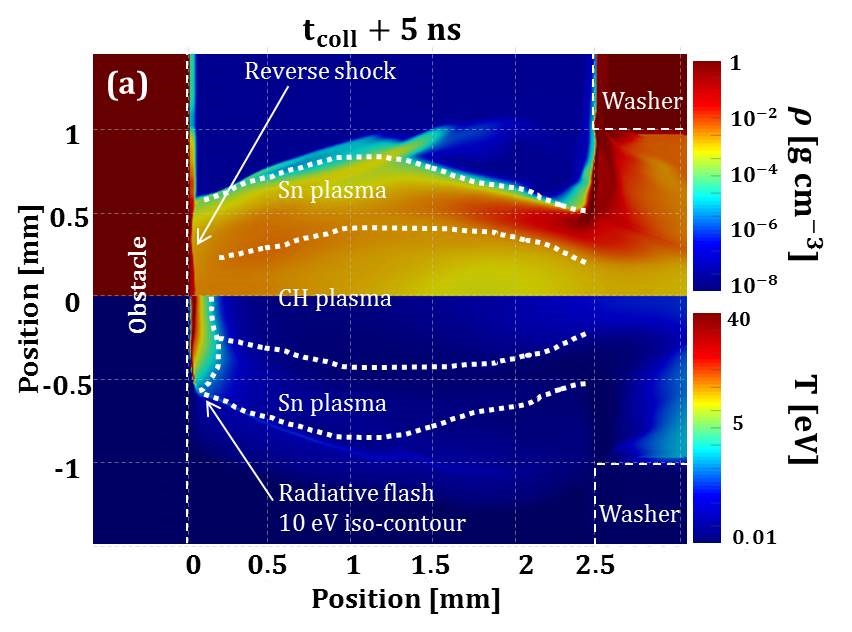}
\includegraphics[width=8.5cm]{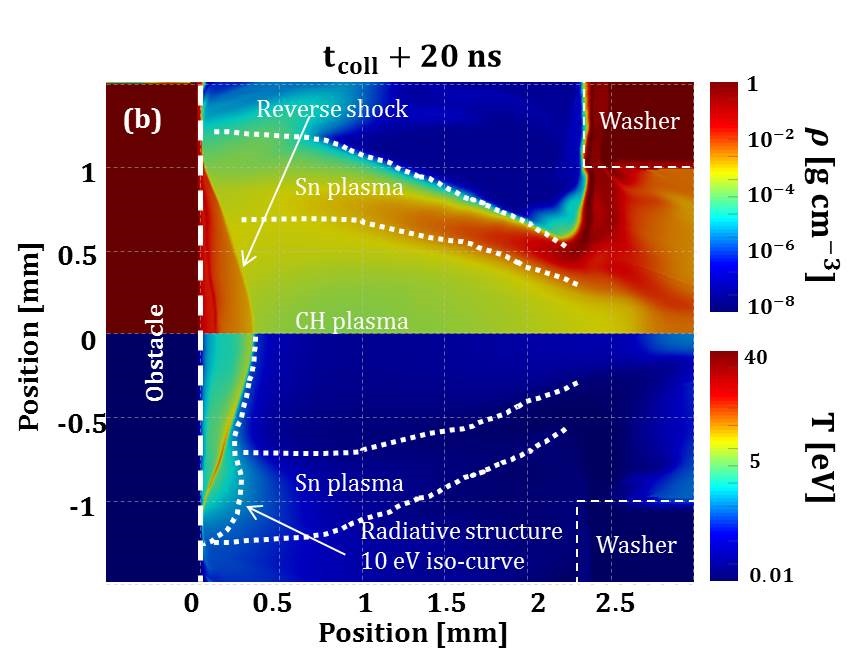}
\caption{Density and temperature maps of the incident flow extracted from 2D numerical simulation around $5$\,ns (a) and $20$\,ns (b) after the collision. Just after the collision, the reverse shock is not propagated whereas a radiative flash is clearly visible in the impact zone. Around $20$\,ns after the collision, the reverse shock has catched up the radiative structure.}\label{fig_sim2}
\end{figure}

The plasma impacts the obstacle at about $20$\,ns (see Figures \ref{fig_40376_1d} and \ref{fig_40376_1d_sop}) leading to the formation of a reverse shock and a transmitted shock in the obstacle (see Figure \ref{1d_bi}). Based on the Rankine-Hugoniot conditions for a strong shock, the typical post-shock temperature can be expressed as a function of the incoming flow velocity, $v_{\text{flow}}$, and the material characteristics (the atomic number $A$ and the charge number $Z$): $ T_{\text{ps}} \propto [{A}/{(Z+1)}][v_{\text{flow}}]^{2}$. Consequently the post-shock temperature is about $T_{ps} \sim 40$\,eV. When the flow impacts the obstacle, a radiative flash (heat wave) is generated and observed both in simulations (see Figure\,\ref{fig_sim2}a) and SOP diagnostics (see Figure \ref{fig_40376_1d_sop} at $20$\,ns). The spatial extension of this emission structure is proportional to the Rosseland mean free path of the pre-shock matter ($L_{R}\propto\lambda_{R}(\rho_{\text{flow}},T_{\text{flow}})$). This leads to the creation of a radiative structure upstream from the shock.

From the 2D simulation, we can determine that the reverse shock propagates into the Sn incoming flow at a slow velocity ($v_{\text{shock}}\sim 5$\,km\,s$^{-1}$) due to the density increase in the post-shock region. With only optical diagnostics, it is not possible to exhibit the position of the reverse shock on experimental data. To probe its structure and its position, X-ray radiographies are necessary. However the visible emitting region in SOP figures proves that the radiative structure is sustained by a reverse shock as shown by the 2D simulations (see Figure\,\ref{fig_40376_1d_sop}). Indeed without the presence of the reverse shock, the strong emission generated at the impact would decrease rapidly which it is not the case here. We define the extension of the radiative structure as the heat zone higher than about $10$ eV. The position of the reverse shock (white line) and the radiative structure (iso-curve at $10$\,eV in black line) determined from simulations are compared with experimental data extracted from the 1D SOP diagnostic in Figure \ref{fig_40376_1d_sop}. The simulation can reproduce the dynamics of the emission region. At the CH arrival at about $10$\,ns after the impact, the reverse shock accelerates at $25$\,km\,s$^{-1}$. The radiative structure persists until the reverse shock catches up with it, which confirms that this particular structure is not a radiative precursor. After $20$\,ns after the collision, the emission region and the post-shock region are mixed (see Figure \ref{fig_sim2}b).

The reverse shock structure depends strongly on the laser parameters and the distance to the obstacle. This is due to the fact that the temperatures of the accretion flow and the post-shock region increase with the laser energy, which strengthen the radiative effects. However the distance of the obstacle decreases the density and the temperature both of the expanding plasma and the post-shock structure because of the longer relaxation in the vacuum. 

\section{Similarity properties}

\begin{table*}
\centering
\begin{tabular}{rccccc}
\hline
Obs. distance [mm] & $2.5$ 	& $2.5$ 	& $3.5$	& $3.5$ 	&  AM  \\
Laser energy [J] & $357$ 	& $1100$	& $329$ 		& $925$ 	& Herculis  \\
\hline
$v_{\text{flow}}$ [km\,s$^{-1}$] & $90$ & $155$ & $95$ & $130$  & $\sim 5000$   \\
$\rho_{\text{ps}}$ [g\,cm$^{-3}$] & $10^{-3}$ & $10^{-2}$ & $10^{-2}$ & $10^{-2}$ & $\sim 10^{-8}$    \\
$T_{\text{ps}}$ [eV] & $25$ & $40$ & $20$ & $30$  & $\sim 10^{4}$   \\
$M$ & $6$ & $10$ & $5$ & $7$  & $\sim 50$  \\
$\chi$ & $0.8$ & $0.1$ & $0.6$ & $0.2$  & $\sim 10^{-2}$  \\
\hline 
\end{tabular}
\caption{Similarity properties of four typical shots at different laser energies and different distances from the obstacle. Values are extracted from the 2D simulations performed with the FCI2 code.}\label{tab_sim_bi}
\end{table*}

The similarity properties are presented in the Table \ref{tab_sim_bi}. In order to evaluate the hydrodynamic parameters of the post-shock region, the quantities are estimated when the reverse shock is located at $200$\,$\mu$m from the obstacle according to the simulations. The characteristics parameters (plasma flow velocity $v_{\text{flow}}$, post-shock density $\rho_{\text{ps}}$ and temperature $T_{\text{ps}}$) and the cooling parameter are compared with the data from the polar prototype AM Herculis star in high-state accretion \citep{Gansicke1995}. The Mach number ($M$) of the reverse shock is defined as $(v_{\text{flow}}-v_{\text{shock}})/c_{s,ps}$. The cooling parameter ($\chi$) in the post-shock region characterizes the balance between the radiative processes and the hydrodynamics in the radiative zone \citep{Falize2011_scaling}, and then the astrophysical relevance of such laboratory targets. In such a process where radiation effects are of some relevance, the cooling parameter offers a qualitative hydrodynamic scaling. However the atomic physics processes might only be partially addressed with such a macroscopic dimensionless number. The cooling parameter is defined by the ratio between the cooling time to the dynamical one in the post-shock region. The dynamical time is defined as the ratio of the length of the post-shock region divided by the sound velocity. The cooling time is determined as the ratio of the internal energy density and the emissivity of the medium. Due to the relatively low temperature and high density of the laboratory plasma compared to the astrophysical regime, the plasma emissivity is not due to the bremsstrahlung cooling. Thus the laboratory emissivity of the medium is given by $\epsilon \sim \kappa_{P}\sigma T^{4}$ where $\kappa_{P}$ and $\sigma$ are respectively the Planck mean opacity and the Stefan-Boltzmann constant.

\begin{figure}[h!]
\centering
\includegraphics[width=8.5cm]{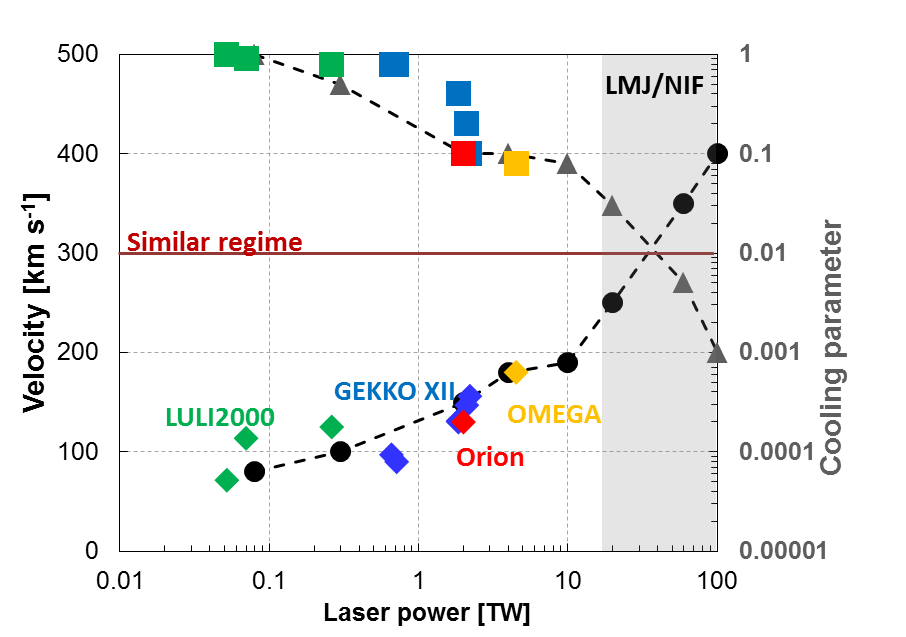}
\caption{Velocities and cooling parameters in POLAR project as a function of the laser power. Simulations are presented with the dotted black lines. Experimental results obtained with intermediate laser facilities are displayed in colour dots. The results obtained with GEKKO XII are presented in blue dots.}\label{velocities}
\end{figure}

The targets achieve a cooling parameter below one, $\chi\sim 0.1-1$ which implies that the cooling losses start to play a role in the structure of the post-shock region. The increase of the obstacle distance decreases the density and the temperature of the expanding flow which cools the post-shock region as shown in Section \ref{section_simu}. Consequently the increase of the obstacle distance increases the cooling parameter. By increasing the laser energy, the cooling parameter can reach $0.1$. In these laboratory conditions, although the cooling parameter is below one, the radiative losses play a role in the evolution of the plasma but do not dominate the dynamics as in the astrophysical regime ($\chi\lesssim10^{-2}$). The GEKKO XII facility allows us to achieve a radiative regime similar to those achieved on the LULI2000 facility\citep{Falize2011_POLAR,Falize2012} for low-energy configuration ($\chi\sim1$) and those achieved on the Orion facility\citep{Cross2016} at high-energy configuration ($\chi\sim0.1$) (see Figure \ref{velocities}). The intermediate energies allow us to reach new radiative hydrodynamic regimes ($\chi\sim 0.1-0.5$), not studied yet. The velocities, associated to the range of laser energies used on GEKKO XII facility, measured with the SOP and the shadowgraphy diagnostics, vary between $90-160$\,km\,s$^{-1}$ for the different types of target designs. The obtained radiative regime links and it fills the gap between the LULI2000 and Orion regimes as shown in Figure \ref{velocities}. In order to achieve similar regime at the laboratory scale, the scaling laws point out that the incident flow must reach about $300$\,km\,s$^{-1}$ and the post-shock medium must be dominated by the bremsstrahlung losses with $\chi \sim 10^{-2}$. These conditions are illustrated by the horizontal line in Figure \ref{velocities}. Targets used on intermediate laser facilities, such as the GEKKO XII laser facility, cannot reach the astrophysical similar regime. Thus, these experimental regimes should be accessible in megajoule laser facilities such as the Laser Megajoule (LMJ) and the National Ignition Facility (NIF).

\section{Conclusion} 

The experimental results obtained during this GEKKO XII experiment allows us to study new experimental design without tube. We characterize the incident flow and the radiation hydrodynamic regime of the produced reverse shock structure. Experimental data were compared to 2D numerical simulation in the GEKKO XII laser conditions. Despite its radial expansion due to the absence of the tube, the plasma flow is well collimated in vacuum by a nozzle-like structure. After the impact, a strong radiative flash is created into the CH incoming flow due to the Rosseland mean free path discontinuity between the Sn and the CH materials. The propagation of this wave is powered by the reverse shock and its extension is proportional to the plastic Rosseland mean free path. This strong emission region is visible in SOP diagnostics in agreement with forecast of numerical simulations. The radiative structure is sustained by a reverse shock structure which can not be experimentally exhibited because of the lack of X-ray radiographies. Thanks to the flexibility on the laser energy produced on GEKKO XII, the radiative losses in the post-shock region can start playing a role in the evolution of the plasma but do not dominate the dynamics as in the astrophysical regime. The achieved radiative regimes are between those obtained with previous POLAR experiments in the LULI2000 facility and those achieved on the Orion facility. The good compatibility between all diagnostics and the simulations demonstrates that the experiment successfully catches the dynamics of the expanding flow and the post-shock region. 

This new target will allow us to integrate an external magnetic field and study magnetic collimation of the accreting flow \citep{Albertazzi2018}. Using a magnetic collimation could open new ways to generate more relevant experiments by limiting the external hydrodynamical effects in laboratory and by increasing the comparison between laboratory and astrophysical scales.  

\section*{Acknowledgements}
We acknowledge the GEKKO XII laser facility staff for their valuable support. Part of this work was supported by the "Programme National de
Physique Stellaire" (PNPS) of CNRS/INSU, France. This work was also partly supported by the ANR Blanc grant no. 12-BS09-025-01 SILAMPA and the LABEX Plas@par grant no. 11-IDEX-0004-02 from the French agency ANR. The authors would like to thank E. Lefebvre for several fruitful discussions.


\begin{thebibliography}{00}
\bibitem[Frank et al.(2002)]{Franck2002} Frank, J., King, A., \& Raine, D.~J.\ 2002, Accretion Power in Astrophysics, Cambridge University Press, February
\bibitem[Maoz et al.(2014)]{Maoz2014} Maoz, D., Mannucci, F., \& Nelemans, G.\ 2014, ARAA, 52, 107 
\bibitem[Riess et al.(1998)]{Riess1998} Riess, A.~G., Filippenko, A.~V., Challis, P., et al.\ 1998, Astrophys. J., 116, 1009 
\bibitem[Warner(1995)]{Warner1995} Warner, B.\ 1995, Cambridge Astrophysics Series, 28,
\bibitem[Cropper(1990)]{Cropper1990} Cropper, M.\ 1990, Space Sci. Rev., 54, 195 
\bibitem[Wu(2000)]{Wu2000} Wu, K.\ 2000, Space Science Rev., 93, 611 
\bibitem[Ramsay \& Cropper(2004)]{Ramsay2004} Ramsay, G., \& Cropper, M.\ 2004, Not. R. Astron. Soc., 347, 497
\bibitem[Bonnet-Bidaud et al.(2015)]{Bonnet2015} Bonnet-Bidaud, J.~M., Mouchet, M., Busschaert, C., et al. \ 2015, Astron. Astrophys., 579, A24 
\bibitem[Mouchet et al.(2017)]{Mouchet2017} Mouchet, M., Bonnet-Bidaud, J.~M., Van Box Som, L., et al. \ 2017, Astron. Astrophys., 600, A53 
\bibitem[Van Box Som et al.(2018)]{VBS2018} Van Box Som, L. Falize, E., Bonnet-Bidaud, J.~M., et al. \ 2018, Mon. Not. R. Astron. Soc., 473, 3158
\bibitem[Falize et al.(2009)]{Falize2009_scaling} Falize, {\'E}., Bouquet, S., \& Michaut, C.\ 2009, Astrophys. Space Sci., 322, 107 
\bibitem[Falize et al.(2011)]{Falize2011_scaling} Falize, {\'E}., Michaut, C., \& Bouquet, S.\ 2011, Astrophys. J., 730, 96
\bibitem[Falize et al.(2011)]{Falize2011_2T} Falize, {\'E}., Dizi{\`e}re, A., \& Loupias, B.\ 2011, Astrophys. Space Sci., 336, 201 
\bibitem[Ryutov et al.(2000)]{Ryutov2000} Ryutov, D.~D., Drake, R.~P., \& Remington, B.~A.\ 2000, Astrophys. J. Series, 127, 465 
\bibitem[Korneev et al. (2015)]{Korneev2015}  Korneev, Ph., d'Humi{\`e}res, E., \& Tikhonchuk, V.\ 2015, Phys. Rev. E, 91, 043107 
\bibitem[Falize et al.(2011)]{Falize2011_POLAR} Falize, {\'E}., Loupias, B., Ravasio, A., et al.\ 2011, Astrophys. Space Sci., 336, 81 
\bibitem[Falize et al.(2012)]{Falize2012} Falize, {\'E}., Ravasio, A., Loupias, B., et al.\ 2012, High Energy Density Physics, 8, 1 
\bibitem[Cross et al.(2012)]{Cross2016} Cross, J.~E., Gregori, G., Foster, J.~M., et al.\ 2016, Nature Com., 7, 11899
\bibitem[Krauland et al.(2013)]{Krauland2013} Krauland, C.~M., Drake, R.~P., Kuranz, C.~C., et al.\ 2013, Phys. Plasmas 20, 056502
\bibitem[Krauland et al.(2013)]{Krauland2013b} Krauland, C.~M., Drake, R.~P., Kuranz, C.~C., et al.\ 2013, Astrophys. J. Lett., 762, L2 
\bibitem[Atzeni et al.(2013)]{Atzeni} Atzeni, S. \& Meyer-ter-Vehn, J. \ 2004,  The physics of inertial fusion. Beam plasma interaction,
hydrodynamics, hot dense matter, Oxford University Press
\bibitem[Schurtz et al.(2000)]{Schurtz2000} Schurtz, G.~P., Nicola{\"i}, P.~D., \& Busquet, M.\ 2000, Physics of Plasmas, 7, 4238 
\bibitem[Busschaert et al.(2013)]{Busschaert2013} Busschaert, C., Falize, {\'E}., Loupias, B., et al.\ 2013, New Journal of Physics, 15, 035020
\bibitem[G{\"a}nsicke et al.(1995)]{Gansicke1995} G{\"a}nsicke, B.~T., Beuermann, K., \& de Martino, D.\ 1995, Astron. Astrophys., 303, 127
\bibitem[Albertazzi et al.(2018)]{Albertazzi2018} Albertazzi, B., Falize, \'E., \& Yurchak, R., et al.\ 2018, High Power Laser Science and Engineering, accepted
\end{thebibliography}
\end{document}